# A Non-Phylogenetic Conceptual Network Architecture for Organizing Classes of Material Artifacts into Cultural Lineages

**Liane Gabora (liane.gabora@ubc.ca), Stefan Leijnen, Tomas Veloz (tomas.veloz@ubc.ca) and**
Department of Psychology, University of British Columbia
Okanagan campus, 3333 University Way
Kelowna BC, V1V 1V7, Canada

**Carl Lipo (clipo@csulb.edu)**
Department of Anthropology and IIRMES, California State University, Long Beach
1250 Bellflower Boulevard
Long Beach, CA 90840, USA

**Abstract**

The application of phylogenetic techniques to the documentation of cultural history can present a distorted picture due to horizontal transmission and blending. Moreover, the units of cultural transmission must be communicable concepts, rather than conveniently measurable attributes, and relatedness between elements of culture often resides at the conceptual level, something not captured by phylogenetic methods, which focus on measurable attributes. (For example, mortars and pestles are as related as two artifacts could be, despite little similarity at the attribute level.) This paper introduces a new, cognitively inspired framework for chronicling material cultural history, building on Lipo's (2005) network-based computational approach. We show that by incorporating not just superficial attributes of artifact samples (e.g. fluting) but also conceptual knowledge (e.g. information about function), a different pattern of cultural ancestry emerges.

## Introduction

The efforts of biologists, phylogeneticists, and others, have culminated in an impressively detailed understanding of how the living things of today evolved. We can trace the ancestral origins of our eyes and fingers, and even certain behavioral traits such as mating preferences. However, we lack comprehensive knowledge of patterns of relatedness of elements of culture, even restricting ourselves just to material artifacts.

The paper discusses difficulties that have arisen attempting to chronicle material cultural history using phylogenetic and network based approaches. We then describe our new conceptual network approach. The insight that guides this approach is: since artifacts are the product of minds that encode representations of them not just at the attribute level but also at an abstract, conceptual level, to reconstruct material cultural evolution it is necessary to incorporate how artifacts are conceived, and how these conceptions interact in a human mind. We introduce a computer program that is able to construct such networks from both attribute data and conceptual information.

## Phylogenetic Approaches

Since artifacts undergo 'descent with modification', the theory of natural selection appears to offer a means for explaining cultural history. Accordingly, phylogenetic methods such as cladistics are routinely borrowed from biology and applied in an archaeological context (O'Brien & Lyman 2003; O'Brien, Darwent & Lyman, 2001). In cladistic representations of archaeological data, the measured attributes of a 'taxon' of artifact are listed as a number string. The position in the string is loosely analogous to the concept of gene, and the number at that position is loosely analogous to the concept of allele. Thus if a taxon is represented by 132 then the first attribute is in state one, the second is in state three, and the third is in state two. For example, consider the representation of early projectile points from the Southeastern United States shown in Figure 1 (O'Brien et al., 2001). The data consist of metric and morphological measurements with respect to eight attributes, each of which can take from two to six possible states. Thus for example if fluting is absent in a particular artifact it has a 1 in position VII, and if fluting is present it has a 2. Seventeen 'taxa' are identified, and the pattern is such that one common ancestor (identified as KDR) gave rise to sequential branchings that culminated in 16 different taxa. This technique provides an intuitively meaningful (although potentially misleading) means of capturing structural change. The 'root taxon' at the far left is the most primitive, and early branch points represent changes that provided the structural constraints that shaped more recent changes. For example, much as evolution of the backbone paved the way for limbs, evolution of containers paved the way for spouts and handles.

Phylogenetic approaches have also been applied to culture in more complex ways. For example, relationships amongst different elements of culture have been analyzed by comparing their phylogenetic trees (Holden & Mace, 2003). The procedure involves running a series of forward models, one in which the phenomena are assumed to evolve completely independently, another in which one kind of correlation is assumed (*e.g.* matriliny with cattle), another in which a different correlation is assumed (*e.g.* patriliny with cattle). These are compared to the language phylogeny, which is assumed to be the most accurate available cultural history tree, to determine which gives the best match. This method can indeed unearth relationships amongst different elements of culture. It was found, for example, that the spread of pastoralism in Sub-Saharan Africa is associated with a shift from matriliny to patriliny. However the method is ineffective if there is rampant blending of cultural elements, and it does not generate information about why or how elements of culture are related.

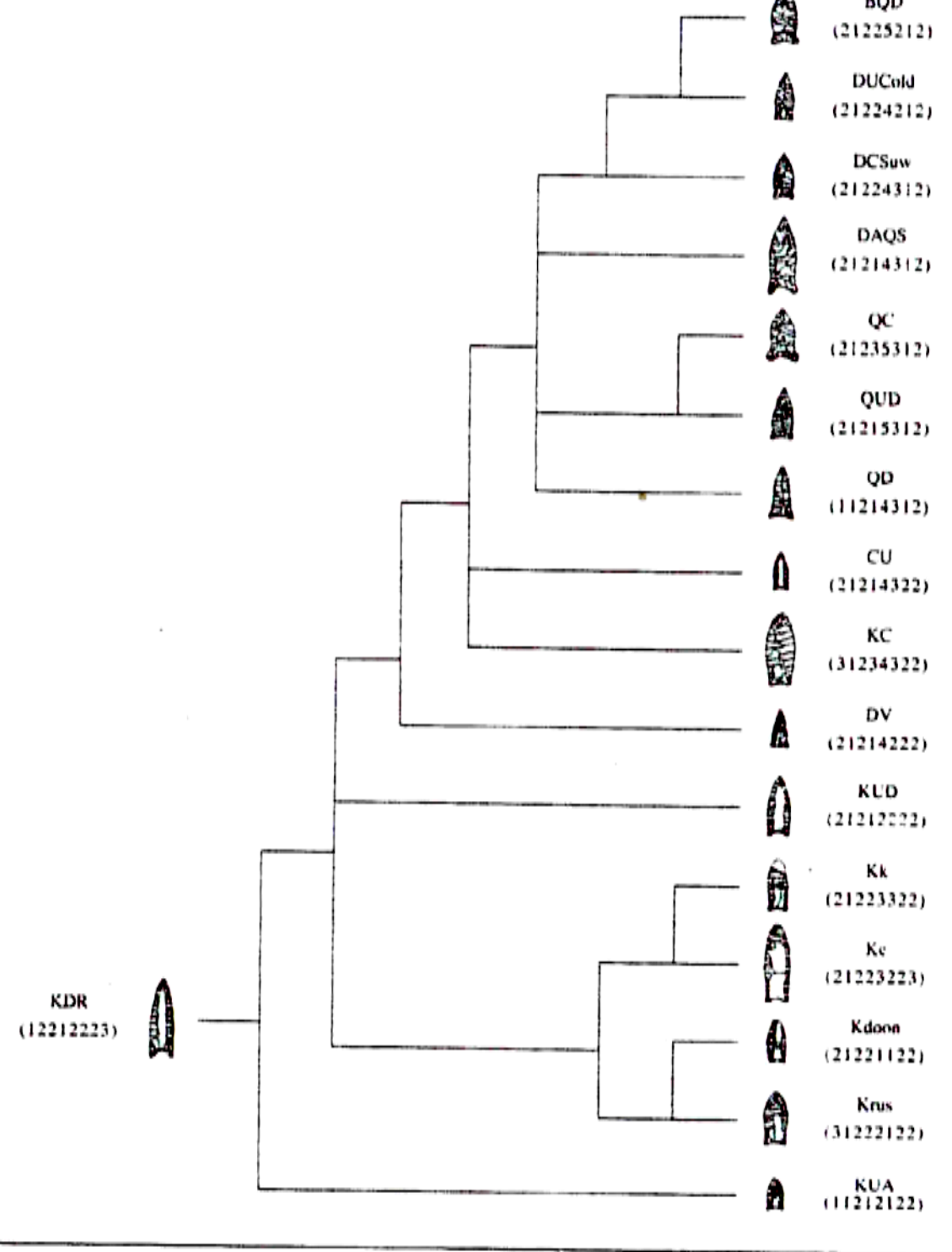

Figure 1. Phylogenetic representation of PaleoIndian period projectile points from the Southeastern United States with 17 taxa defined by 18 attributes. From O'Brien et al., 2001.

## Shortcomings of Phylogenetic Approaches

Despite the intuitiveness and scientific rigor of phylogenetic/cladistic approaches, and some apparent successes applying them to culture, concerns have been raised about distortions generated by these cultural applications (Gabora, 2006; Lipo, 2005; Tëmkin & Eldredge, 2007; Terrell, Hunt, & Gosden, 1997). We now examine these concerns.

**Similarity Need Not Reflect Homology**. Phylogenetic methods assume that similarity reflects homology, i.e. that two species are similar because they are related. Specifically, it assumes that, either (1) one is descended from the other, in which case shared traits were transmitted *vertically*, or (2) they are descended from a *common ancestor,* which is depicted as a branch point. For example, common ancestry can occur through *fission*, in which a population splits in two, which become increasingly differentiated.

However, similarity need not reflect homology. Artifacts may arise independently yet be similar because they are alternative solutions within similar design constraints. Convergent evolution occurs in a biological context too. However, because organisms must solve *many* problems (reproduction, locomotion, digestion, etc.) the probability that a species is mis-categorized on the basis of how it solves any one problem is low. Artifacts, on the other hand, are generally constructed with a single use in mind. (Though artifacts developed for use in one context may be used to solve other problems, e.g., a screwdriver may be used to open a can of paint). Therefore, the probability of mis-categorization arising through the assumption that similarity reflects homology is problematic for artifacts.

**Blending**. Cultural relatedness frequently arises through not just vertical transmission but horizontal (inter-lineage) transmission, which can result in the blending of knowledge from different sources. Since inter-lineage transfer of information is relatively rare in animals, phylogenetic methods are ill-equipped to deal with it. Extensive horizontal transmission gives a bushy, reticulated appearance to a phylogenetic tree, which is misleading because it implies not just chronology but ancestry.

Blending is problematic for cladistic methods because it forces one to parse the data according to predefined attributes or characters. So one is *a priori* discouraged from incorporating data that does not fit into this parsing. In biology, such parsing arises naturally stemming from how traits are genetically encoded. The chosen attributes are characteristic of that species, and the rarity of inter-species mating ensures that they don't change drastically. However, in culture, nothing is *a priori* prohibited from 'mating with' anything else. Those who apply phylogenetics to culture respond that such problems rarely arise in the study of prehistory. On the basis of a set of studies of virtually indistinguishable artifacts, Collard et al. (2006) insinuate that cultural blending is not widely present. This, however, reflects their highly limited choice of artifacts; a brief examination of the contents of any modern house would lead one to a different conclusion. Moreover, even if one is more interested in prehistoric culture than contemporary culture, one seeks not a bag of tricks for assessing relatedness each of which is applicable to certain data sets, but an explanatory framework that fits them all.

**Lack of Objective Measure of Relatedness**. A more fundamental problem with phylogenetic approaches to culture is that they assume it is possible to accurately *measure* the relatedness of artifacts. Whether or not two organisms share a common ancestor is clear-cut; they either are or are not descendents of a particular individual. One can objectively measure what percentage of the genomes of two species overlap, and make conclusions about their degree of genetic relatedness. But in a cultural context, whether or not two artifacts "share a common ancestor" can be arbitrary, and moreover, what is measured is not necessarily what was culturally transmitted.

**Predefined Attributes**. The data of Figure 1 are typical of those to which a phylogenetic approach is amenable because the taxa are very similar to one another. That is, each taxon has one version or another of the considered attributes; there are no major modifications in this lineage. A problem pointed out by Alex Bentley (pers. com.) is that the units considered are those that are most amenable to analysis rather than those that were most likely to have been transmitted from teacher to apprentice. Thus the method documents readily measurable change, not the actual cultural ancestry of the artifact.

## Lipo’s Network (LN) Approach

Network-based methods appear to avert the above problems by simply ordering data according to similarity without necessarily implying common ancestry (Lipo, 2005). Analysis of the same data yields quite a different pattern of evolutionary change. Following O’Brien, samples that are rated the same with respect to all considered attributes are categories together as a particular taxon. Attributes are encoded as a number string. Each position in the string refers to a particular attribute, and the number at a position refers to the state of that attribute for the taxon. This is shown in Figure 2.

Taxa are simply arranged according to the number of attributes by which they differ. The majority of taxa have two lines coming from them, one to a taxon that preceded it, and one to a taxon that followed it; the network does not specify which is which. Those that have more (e.g. 31222122) reflect the existence of multiple other taxa with the same number of differences.

Several aspect of the procedure are noteworthy. First, the network-based approach does not make *a priori* assumptions about the sources of diversity. It is uncommitted with respect to whether differences reflect branching due to fission or blending due to transmission. Second, the method is also uncommitted with respect to chronology. Additional data indicate the directionality of the evolutionary pathway, as shown in Figure 3.

### Limitations of the LN Approach

We believe that in order to avoid the limitations of phylogenetic methods a move in the direction of network representations is inevitable. However, this initial implementation has limitations.

**Considers Only Superficial Attributes.** This approach is suitable for artifacts that are highly similar at the superficial attribute level. However, it cannot to handle artifacts whose similarity resides at the conceptual level. For example, it seems reasonable to hypothesize that the stoplight has (at least) two cultural ancestors—the streetlight and the car--the first contributing the necessary expertise (mastery over the technological design space of external lighting), and the second contributing the necessary motive (control traffic). The second is as crucial as the first; if cars (or something like them) had not come into existence, stoplights would not have come into existence. However, the network approach does not provide a way to document this. Their lack of low-level similarity means that this relationship cannot be reconstructed using this method.

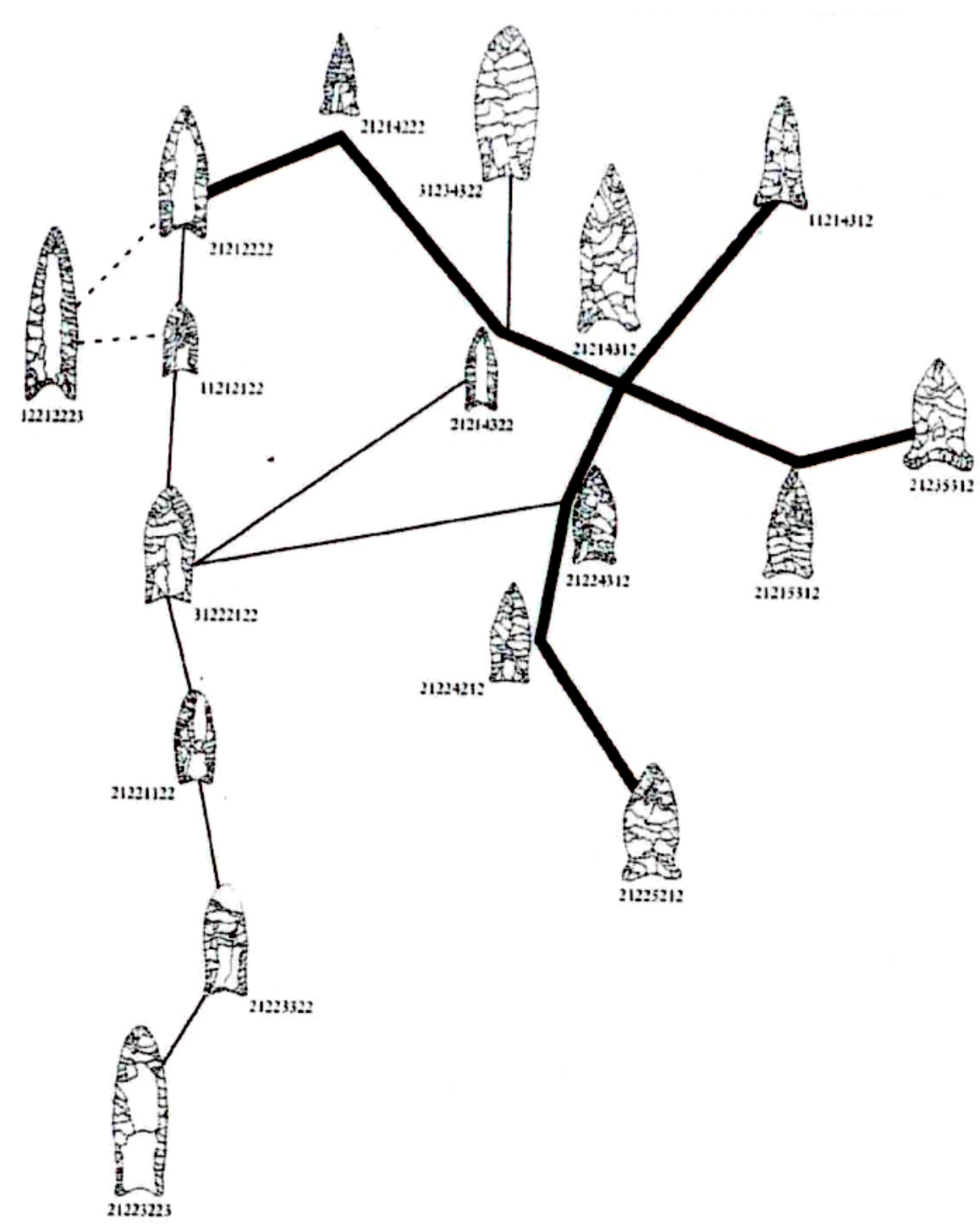

Figure 2. Graph produced by linking taxa to their most similar neighbors. Bold lines represent differences of only one attribute. Thin solid lines show differences of two attributes. Dotted lines show differences of three attributes. The multiple lines connecting taxon 31222122 to other taxa indicate ambiguity due to equivalent number of differences between multiple taxa. From Lipo, 2005.

**Assumes Single-Attribute Change**. The LN architecture assumes that the evolutionary path cannot be resolved when there are multiple attribute differences between neighboring taxa. This is not the case when conceptual structure is taken into account; multiple differences (or even complete lack of similarity) at the attribute level may reflect single changes at the concept level. Moreover, once the conceptual level is introduced, it is no longer necessary to restrict oneself to independent attributes. Indeed, dependencies amongst attributes may indicate the presence of conceptual structure that may hold the clue to the artifact’s evolutionary story.

**Constraints on Attributes.** Third, the length of the number string and the attributes considered are determined *a priori* according to certain rules: attributes must be *independent*, and there must no significant difference in the fitness of alternative states, i.e. only *neutral variation* is considered. The rationale behind these rules is that they rule out similarity due to convergence (e.g. structural constraints). There is also an implied preference for data with taxa that differ from one another by only one attribute because in such cases the pattern of ancestry can be resolved without ambiguity. When there are differences of multiple attributes between a taxon and its nearest neighbor, the evolutionary path cannot be resolved (e.g. the transition from 111 to 122 could occur by way of either 112 or 121). The underlying assumption is that innovation involves one superficial attribute at a time, so a lack of single-attribute change between neighboring classes is assumed to indicate an incomplete data set. However, this assumption is not always met. For example, Tëmkin & Eldredge's (2007) cornet data exhibits "well-documented temporally spaced sequences of "missing links" that likely indicate an actual pattern of ancestry and descent" (p. 150).

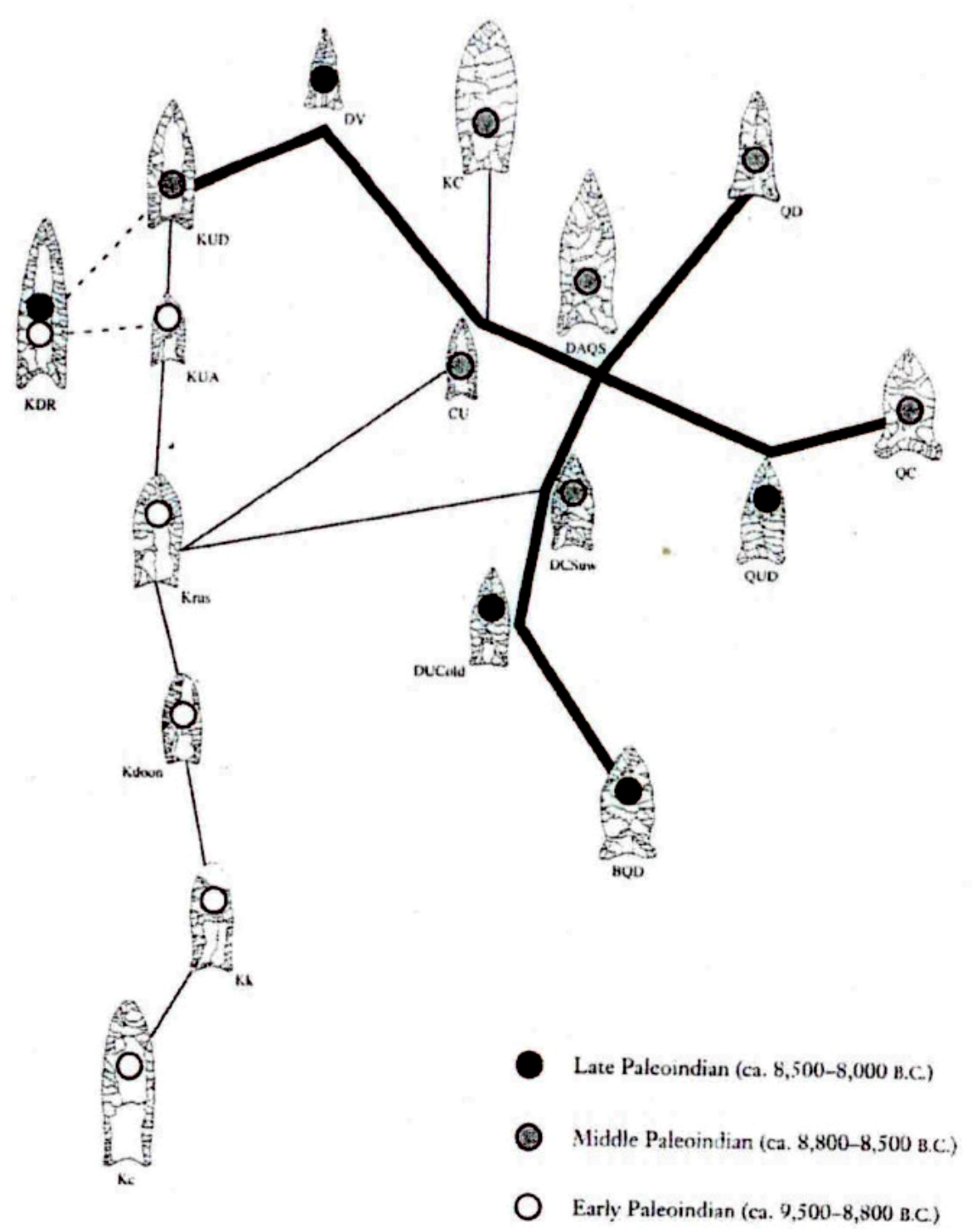

Figure 3. Graphical analysis of projectile point data with temporal information (from Anderson *et al*., 1996) indicated by degree of shading of circles. (From Lipo, 2005).

The network method has the limitation that to chronicle the evolution of a lineage that is increasing in complexity, one would either have to go backwards and add placeholders for traits that did not previously exist, or clump together a great variety of taxa as indistinguishable instances of the terminus. To document the history of human material culture, our framework must accommodate, for example, that this lineage, or one like it, eventually gave rise to the gun. The gun has few of the attributes considered thus far in analyses of this lineage such as 'fluting' or 'arc-shaped base'. Its similarity, indeed our sense that it belongs in this lineage, is conceptual; it reflects the way it is conceived of and used by humans.

In sum, the network method is a sensible, rigorous way of organizing archaeological data. However, due to its assumed independence of attributes, consideration of only superficial attributes, and fixed-length attribute strings, the resulting framework for cultural evolution is fragmentary, limited in application to what many would find the least interesting, or at any rate the least innovative, periods of cultural change.

## The Conceptual Network (CN) Approach

The project described here builds on Lipo's network-based method but adds conceptual structure. As is conventional, concepts are indicated with capitals. Thus an instance of a projectile point is written as 'projectile point' but the concept of one is written as PROJECTILE POINT. The more superficial level of conceptual structure consists of what Rosch (1976) refers to as *basic level concepts* such as PROJECTILE POINT and KNIFE. Basic level concepts mirror the attributes of objects in the external world. This basic level is the level at which items are first perceived, and it is the level at which we generally refer to and interact with them. In some cases it may be more natural to work at a finer level of discrimination and thus consider a more subordinate conceptual level, e.g. BEVELED KNIFE instead of KNIFE. The important thing is that this superficial level be rich in attributes. The less superficial, more abstract level of conceptual structure consists of *superordinate concepts* such as WEAPON. Superordinate concepts often refer to multiple basic level categories (e.g. PROJECTILE POINT and KNIFE are both instances of WEAPON), and they are more general than the level at which we refer to and interact with items (e.g. different kinds of weapon are interacted with in different ways). Basic level concepts and superordinate concepts can take us a long way toward a representation of how objects in the world and their interrelations are conceptualized.

To organize material culture in a way that allows for projectile points to evolve into guns, we incorporate a minimal amount of conceptual structure. The structure of the concept PROJECTILE POINT may include not just that it has certain attributes but also that it is an instance of the concept WEAPON. Sometimes the structure of concepts derives from their history (how they were conceived in the past), and sometimes from other sources (e.g. horizontal transmission or copying error). The cognitive approach uses networks to represent, not just taxa of artifacts, but relationships amongst them as they are conceived of in the mind of a particular population of individuals at a particular time and place.

The program was developed using the object-oriented Java platform with extension packages for working with networks (JUNG) and Excel files (SX). The tool collects meta-data for a set of known samples by asking the user questions about their presumed function and use. The questions are generated using a *conceptual network* that determines which questions are relevant for the sample. This leads to the creation of two networks: an attribute-level only one, and one that incorporates meta-data. Other software functions allow the user to export and import data sets for later use, storing meta-data and networks.

### Data Samples

Data can be entered into the program either manually, filling out fields for each sample, or as a batch excel files that contain all samples to be evaluated. Both import methods require a series of entry fields to be filled out in order for the program to query the user in the next stage. These entries are as follows:

| |
|---|
| **Sample name:** Unique name that identifies the sample<br>*Example:* "Graham3" |
| **Sample attributes:** Features encoded as a numeric string.<br>*Example:* "2262233212221" |
| **Generic Type:** Group to which this sample belongs<br>*Example:* "Graham Cave" |
| **Period:** Estimated period of the sample's original use<br>*Example:* "7,000 – 5,500 B.C." |
| **Location:** Describes where the sample has been found<br>*Example:* "Cooper Site" |
| **Image:** Picture of sample<br>*Example:* Figure 4. |

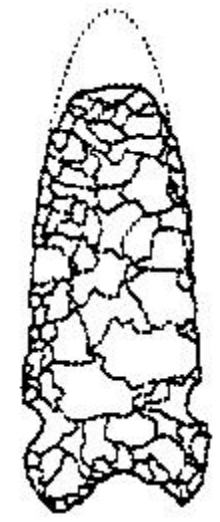

Figure 4. Image of Graham3 sample.

### Conceptual Networks

The samples are described by a set of superficial attributes related to their relative sizes and shapes, the material from which they were constructed, and so forth. Since the intended function of an artifact does not follow unambiguously from these attributes, a human expert capable of deducing function from shape, and who may also have knowledge concerning their location and period, provides additional information to aid the computer program in determining how the samples are related. Following Dunnell (1978), we define function in terms of the relationship between an object and its environment, including both natural and artificial aspects. Variability in the physical aspects of objects sometimes reflects function. For example, broad, thick objects have lower performance values than narrow ones for piercing, and objects that interact with air at any velocity are shaped by aerodynamics. Since the number of possible functions that an artifact could have is potentially infinite, the program asks only those questions that are relevant for a particular sample based on assessment of attributes. Since thus far much of the data has consisted of projectile points, all samples trigger the question, 'Was the sample a projectile point', and 'Was the sample thrown'. Other examples of questions asked include, 'Was the sample used for cutting'.

### Database

Answers given by human experts are stored as meta-data in the program. Since for large datasets, an expert may not be able to handle the full set in one session, sets of samples may be imported from and exported to text format files.

### Generation of Lineages

The program analyses the superficial attributes and abstract (e.g. functional) aspects of samples, and uses this information to generate networks that arrange the artifacts according to how similar they are. Thus the network shows how the artifacts are likely to have evolved chronologically. Relative distance between two samples $x$ and $y$ in the original network is determined by the following algorithm:

$$N(x,y) = H(f(x), f(y))$$

Where

$N$ is the distance without abstract concepts

$H$ is the Hamming distance between two encodings

$f(x)$ is the attribute encoding of $x$

For the CN, the algorithm is expanded with a function over the meta-data:

$$M(x,y) = N(x,y) + D(a(x),a(y))$$

Where $D$ is a binary function that indicates whether two attributes are similar (0) or different, and (1) $a(x)$ is a conceptual level attribute of $x$.

## Results

Although the approach has not yet been thoroughly tested, in every test of ten or more samples so far there is at least one difference in the chronological ordering of between the CN approach and the original network approach. For comparative purposes, we began with the same data that was analyzed using the previously described approaches. An example of actual output of the program is given in Figure 5. Since using the entire data set generates output that is crowded and difficult to parse, the figure just shows a subset of the data. The output shows both the original network approach and the CN approach. In the LN approach, shown to the left, for any sample $x$, it is possible that more than one of the other samples is equally similar to $x$, *i.e.* minimizes the Hamming distance (the $N$ function) with respect to $x$. Therefore, using attributes only, there is a large probability of generating the incorrect lineage. If you look to the samples featured on the upper right, it guessed that the terminal sample 'Calfcreek' is most closely related to the topmost sample, 'Graham4'. Indeed based on the superficial attributes only this was a reasonable guess.

In the CN approach, however, using conceptual information (the *M* function) we can distinguish the correct ordering on the basis of higher-level information. We see that the LN approach guessed incorrectly, and that 'Calfcreek' is actually more closely related to 'Graham2', the one below it, than to 'Graham4', the one guessed using the LN.

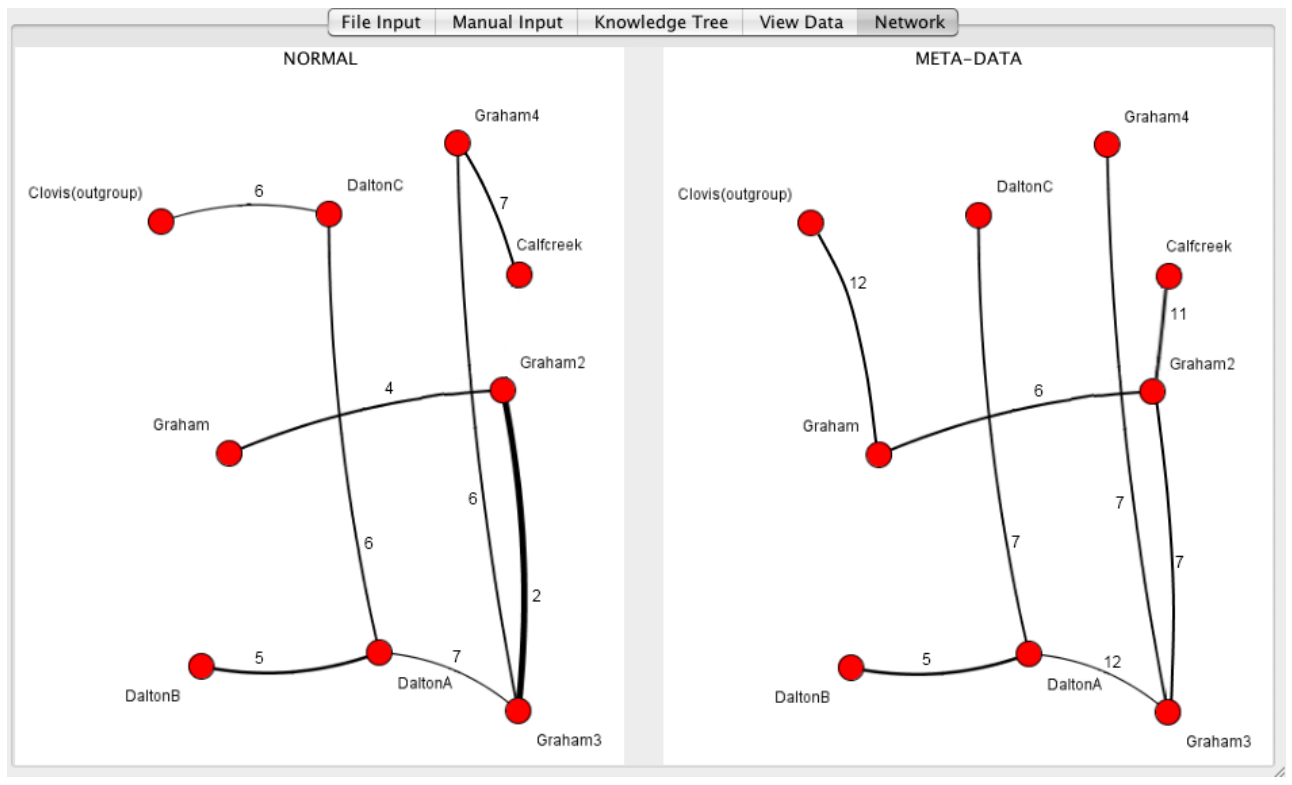

Figure 5. Two examples of network output given the same input data. Circles represent particular samples. Numbered lines give estimates of relatedness (lower numbers more closely related). The output on the left makes use of superficial attributes only. The output on the right additionally makes use of conceptual meta-data.

## Discussion

To reconstruct the history of the objects we build and use requires us to consider conceptual relationship, indeed to reconstruct the history of conceptual change in the minds that created them. The conceptual network approach introduced here avoids inherent in phylogenetic approaches. It builds on an earlier network-based model, by adding the capacity to make use of not just superficial attributes of artifacts but also abstract, knowledge referred to as meta-level data. Though for this initial analysis for comparative purposes we stuck with data that had been previously analyzed using other approaches, the current approach can readily be applied to chronicling of patterns of interrelatedness amongst artifacts of *different* kinds (e.g. one tool might fall into disuse when a superior tool comes into existence, or the tool for procuring a certain food might be expected to appear at the same time and location as the tool for processing it). The approach is in its infancy; we continue to improve the program through application of research from cognitive science on concept combination and the formation of hierarchical conceptual structure (e.g. Coley, Hayes, Lawson, & Moloney, 2004; Kemp & Tenenbaum, 2008). Though preliminary, we believe that the approach holds promise in the quest to understand the ancestry of the multitude of artifacts we have created.

## Acknowledgments

We would like to acknowledge grants to Liane Gabora from the *Social Sciences and Humanities Research Council of Canada* and the Concerted Research Program of the *Fund for Scientific Research of Belgium*.

## References

Anderson, D. G., & Sassaman, K. E. (1996). *The Paleoindian and Early Archaic Southeast.* University of Alabama Press.

Coley, J. D., Hayes, B., Lawson, C., & Moloney, M. (2004). Knowledge, expectations, and inductive inferences within conceptual hierarchies. *Cognition, 90,* 217-253.

Collard, M., Shennan, S. L., Tehrani, J. J. (2006). Branching, blending, and the evolution of cultural similarities and differences among human populations. *Evolution and Human Behavior 27*, 169–184.

Dunnell, R. C. (1978). Archaeological Potential of Anthropological and Scientific Models of Function. In R. C. Dunnell & E. S. Hall (Eds.) *Archaeological Essays in Honor of Irving Benjamin Rouse*. The Hague: Mouton

Gabora, L. (2006). The fate of evolutionary archaeology: Survival or extinction? *World Archaeology 38*(4), 690–696.

Holden, C. J., Mace, R. (2003). Spread of cattle led to the loss of matriliny in Africa: a co-evolutionary analysis. *Proceedings of the Royal Society B: Biological Sciences 270*(1532), 2425-2433. ISSN: 0962-8452

Kemp, C., & Tenenbaum, J. B. (2008). The discovery of structural form. *Proceedings of the National Academy of Sciences. 105*(31), 10687-10692.

Lipo, C. P. (2005). The resolution of cultural phylogenies using graphs. In Lipo, C. P., O'Brien, M. J., Collard, M., & Shena, S. J. (Eds.) *Mapping our Ancestors*. New Brunswick: Transaction Publishers.

O'Brien, M. J., Darwent, J., & Lyman, R. L. (2001). Cladistics is useful for reconstructing archaeological phylogenies: Paleoindian points from the Southeastern United States. *Journal of Archaeological Science 28*, 1115-1136.

O'Brien, M. J., Lyman, R. L., Saab, Y., Saab, E., Darwent, J., & Glover, D. S. (2002). Two issues in archaeological hylogenetics: Taxon construction and outgroup selection. *Journal of Theoretical Biology 215*, 133–150.

Tëmkin, I., & Eldredge, N. (2007). Phylogenetics and material cultural evolution. *Current Anthropology 48*, 146–153.

Terrell, J. E., Hunt, T. L., & Gosden, C. (1997). The dimensions of social life in the Pacific: Human diversity and the myth of the primitive isolate. *Current Anthropology 38*, 155–95.